\documentclass[twoside,12pt]{article}
\usepackage{epsfig}

\newcommand{\be}{\begin{equation}}
\newcommand{\ee}{\end{equation}}
\newcommand{\bea}{\begin{eqnarray}}
\newcommand{\eea}{\end{eqnarray}}

\begin{document}

\title{ \vspace{1cm} Dynamics of strange, charm and high momentum
hadrons in relativistic nucleus-nucleus collisions}
\author{W.\ Cassing$^1$, K.\ Gallmeister$^1$,
E. L.\ Bratkovskaya\thanks{Supported by DFG} $^{2}$,
C.\ Greiner$^2$, and H.\ St\"ocker$^2$  \\ \\
$^1$Institut f\"{u}r Theoretische Physik,
   Universit\"{a}t Giessen,  Germany\\
$^2$Institut f\"{u}r Theoretische Physik,
   Universit\"{a}t Frankfurt,  Germany}
   \date{}
\maketitle
\begin{abstract} We investigate hadron production and attenuation of
hadrons with strange and charm quarks (or antiquarks) as well as
high transverse momentum hadrons in relativistic nucleus-nucleus
collisions from 2 $A\cdot$GeV to 21.3 $A\cdot$TeV within two
independent transport approaches (UrQMD and HSD). Both transport
models are based on quark, diquark, string and hadronic degrees of
freedom, but do not include any explicit phase transition to a
quark-gluon plasma.  From our dynamical calculations we find that
both models do not describe the maximum in the $K^+/\pi^+$ ratio
at 20 - 30 A$\cdot$GeV in central Au+Au collisions found
experimentally, though the excitation functions of strange mesons
are reproduced well in HSD and UrQMD. Furthermore, the transport
calculations show that the charmonium recreation by $D+\bar{D}
\rightarrow J/\Psi + \ meson$ reactions is comparable to the
dissociation by 'comoving' mesons at RHIC energies contrary to SPS
energies. This leads to the final result that the total $J/\Psi$
suppression as a function of centrality at RHIC should be less
than the suppression seen at SPS energies where the 'comover'
dissociation is substantial and the backward channels play no
role. Furthermore, our transport calculations -- in comparison to
experimental data on transverse momentum spectra from $pp$, d+Au
and Au+Au  reactions -- show that pre-hadronic effects are
responsible for both the hardening of the hadron spectra for low
transverse momenta (Cronin effect) as  well as the suppression of
high $p_T$ hadrons. The mutual interactions of formed hadrons are
found to be negligible in central Au+Au collisions at $\sqrt{s}$ =
200 GeV for $p_T \geq$ 6 GeV/c and the sizeable suppression seen
experimentally is attributed to a large extent to the interactions
of 'leading' pre-hadrons with the dense environment.
  \end{abstract}

\section{Introduction}
The dynamics of ultra-relativistic nucleus-nucleus collisions at AGS, SPS
and RHIC energies are of fundamental interest with respect to the
properties of hadronic/partonic systems at high energy densities as
encountered in the early phase of the 'big bang'.  Especially the
formation of a quark-gluon plasma (QGP) and its transition to
interacting hadronic matter has motivated a large community for about
20 to 30 years by now \cite{QM01}. However, even after more than a
decade of experiments at the SPS and in the last years at RHIC the complexity of
the dynamics has not been unraveled and no conclusive evidence has been
obtained for the formation of the QGP and/or the properties of the
phase transition.

Apart from the light and strange flavor
($u,\bar{u},d,\bar{d},s,\bar{s}$) quark physics and their hadronic
bound states in the vacuum ($\pi, K, \phi$ etc.) the interest in
hadronic states with charm flavors ($c, \bar{c}$) has been rising
additionally in line with the development of new experimental
facilities. This relates to the charm production cross section in
$pN$, $\pi N$, $pA$ and $AA$  reactions as well as to their
interactions with baryons and mesons which determine their
properties (spectral functions) in the hadronic medium. Especially
the $J/\Psi$ suppression at SPS energies has been discussed as a
signature for a QGP phase due to Debye screening \cite{Matsui}. On
the other hand, it has been pointed out - within statistical
models - that at RHIC energies the charmonium formation from open
charm + anticharm mesons might become essential
\cite{Rafelski,Goren0} and even exceed the yield from primary $NN$
collisions. Thus charmonium suppression or enhancement is of
central importance for an understanding of nucleus-nucleus
collisions at RHIC energies.

Additionally,  transverse mass (or momentum) spectra of hadrons
are in the center of interest. On the one hand the suppression of
high transverse momentum hadrons is investigated in Au+Au
reactions relative to $pp$ collisions at RHIC energies of
$\sqrt{s} =$ 200 GeV  \cite{survey}, since the propagation of a
fast quark through a hot colored medium (QGP) should be different
than in ordinary high density hadron matter \cite{Wang,Baier}. In
fact, the PHENIX \cite{PHENIX} and STAR \cite{STAR} collaborations
have reported a large relative suppression of hadron spectra for
transverse momenta $p_T$ above $\sim$ 3-4~GeV/c which might point
towards the creation of a QGP, since this suppression is not
observed in d+Au interactions at the same bombarding energy per
nucleon \cite{E1,E2}. But it is not clear presently, to which
extent this suppression might be due to ordinary hadronic final
state interactions \cite{Kai}, too. On the other hand, the
measured transverse mass ($m_T-$) spectra  of kaons at   AGS and
SPS energies  show the opposite behavior for {\it soft} hadrons,
i.e. a substantial {\it hardening} of the spectra in central Au+Au
collisions relative to $pp$ interactions (cf. also  \cite{Goren}).
The {\it flat} spectra observed are commonly attributed to strong
collective flow that is absent in the respective $pp$ or $pA$
data. An open question currently is, if the collective flow seen
experimentally might be attributed to final state interactions of
formed hadrons, too.

In our studies we use two independent transport models that employ
hadronic and string degrees of freedom, i.e. UrQMD \cite{URQMD1,URQMD2}
and HSD \cite{Geiss,Cass99}. They take into account the formation and
multiple rescattering of hadrons and thus dynamically describe the
generation of pressure in the early phase - dominated by strings - and
the hadronic expansion phase. Both transport approaches are matched to
reproduce the nucleon-nucleon, meson-nucleon and meson-meson cross
section data in a wide kinematic range. We point out explicitly, that
no parton-parton scattering
processes are included in the studies below contrary to the multi-phase
transport model (AMPT) \cite{Ko_AMPT}, which is currently employed from
upper SPS to RHIC energies.
Whereas the underlying concepts of UrQMD and HSD are very similar,
the actual realizations differ
considerably.

\section{Strange and charm hadrons}

\subsection{{\it Numerical implementation of 'hard' probes}}
Whereas the dynamics of hadrons with $u,d,s$ (and $\bar{u}, \bar{d}, \bar{s}$) quarks
is described nonperturbatively in HSD and UrQMD, the dynamics of 'rare' species
needs a perturbative modeling in order to achieve acceptable statistics. For the
production and propagation of open charm mesons, charmonia or
hadrons with high transverse momentum ($>$ 2.0 GeV/c) we employ the
perturbative scheme as  used in Refs.
\cite{Cass01a,Cass01,Geiss99,Cass97}. We recall that initial hard processes (such as
$c\bar{c}$, Drell-Yan or high $p_T$ hadron production from $NN$
collisions) are 'precalculated' to achieve a scaling of the
inclusive cross section with the number of projectile and target
nucleons as $A_P \times A_T$ when integrating over impact
parameter (cf. Ref. \cite{Cass01}).

\noindent Each open charm meson, charmonium or high $p_T$ hadron is produced in the
transport calculation with a
weight $W_i$ given by the ratio of the actual production cross
section divided by the inelastic nucleon-nucleon cross section,
i.e.
\begin{equation}
\label{weight}
 W_i = \frac{\sigma_{NN \rightarrow h(p_T) +
x}(\sqrt{s})}{\sigma_{NN}^{inelas.}(\sqrt{s})}.
\end{equation}
We then follow the motion of these hadrons within the full
background of strings/hadrons by propagating them as free
particles, i.e. neglecting in-medium potentials, but compute their
collisional history with baryons and mesons or quarks and
diquarks. For reactions with diquarks we use the corresponding
reaction cross section with baryons multiplied by a factor of 2/3.
For collisions with quarks (antiquarks) we adopt half of the cross
section for collisions with mesons and for the leading quark
(formed) baryon collision a factor of 1/3 is assumed. The final states
in an inelastic collision are
then modeled by the JETSET fragmentation scheme in the same way as
for ordinary hadron-hadron collisions.

\subsection{\it Excitation functions of mesons in central collisions of heavy nuclei}

In order to provide an overview on meson production we show in
Fig. 1 the calculated excitation function of $\pi^+, K^\pm, \eta,
\phi, D, \bar{D}$ and $J/\Psi$ mesons in central $Au + Au$
collisions from lower SIS to top RHIC energies without employing
any self energies for these mesons  in the HSD transport approach
(cf. Ref. \cite{Cass01a,Cass01}). We mention that the excitation
functions of the strange hadrons from HSD and UrQMD match rather
well with the data available, however, the sharp maximum in the
$K^+/\pi^+$ ratio at $\sim$ 20 to 30 A$\cdot$GeV in central Au+Au
(Pb+Pb) collisions is missed \cite{Weber02}. This also holds for
the slope of the $K^\pm$ transverse mass spectra above $\sim$ 5
A$\cdot$GeV in central Au+Au (Pb+Pb) collisions
\cite{Brat03PRL,Horst}.

The $\bar{D}$-mesons with a $\bar{c}$ are produced more frequently
at low energies due to the associated production with $\Lambda_c,
\Sigma_c, \Sigma_C^*$ similar to the kaon case, where kaon+hyperon
production is more frequent than $K+\bar{K}$ production. At roughly
15 A$\cdot$GeV the cross sections for open charm and charmonia are
similar, while the ratio of open charm to charmonium bound states
increases rapidly with energy. Again, this behaviour is comparable
to the excitation functions in the strangeness sector when
comparing $K^+,K^-$ and $\phi$-mesons. Since the excitation
function for open charm drops very fast with decreasing bombarding
energy, experiments at 20 to 30 A$\cdot$GeV at the future GSI
facility will be a challenging task since the multiplicity of the
other mesons is higher by orders of magnitude. On the other hand,
the perspectives for open charm measurements at RHIC appear
promising since about 15 $c\bar{c}$ (or $ D\bar{D})$ pairs should
be created in central $Au + Au$ collisions at $\sqrt{s}$ = 200 GeV
according to the HSD transport calculations.

\begin{figure}[t]
\begin{center}
\begin{minipage}[l]{12 cm}
\epsfig{file=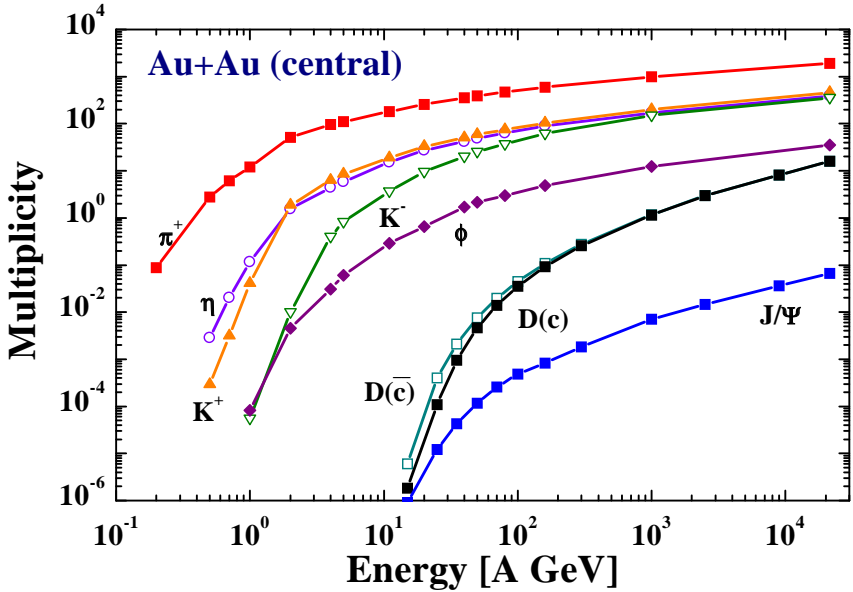,scale=0.75}
\end{minipage}
\phantom{a}\begin{minipage}[l]{5.8cm} \caption{The multiplicities
for $\pi^+, \eta, K^+, K^-$, $\phi$, $D,\bar{D}$ and
$J/\Psi$-mesons for central collisions of $Au+Au$ as a function of bombarding energy.
In the HSD calculations elastic and inelastic reactions of all hadrons are included, but no
in-medium modi\-fications of their spectral functions (The figure is taken from
Ref. \protect\cite{Cass01}).} \label{Fig1}
\end{minipage}
\end{center}
\end{figure}

\subsection{{\it Open charm and charmonia at SPS and RHIC}}
We calculate open charm and charmonium production at
SPS and RHIC energies (within the HSD transport approach
for the overall reaction dynamics) using
parametrizations for the elementary production channels including the
charmed hadrons $D, \bar{D}, D^*, \bar{D}^*, D_s, \bar{D}_s, D_s^*,
\bar{D}_s^*,$ $J/\Psi, \Psi(2S), \chi_{2c}$ from $NN$ and $\pi N$
collisions. The latter parametrizations are fitted to PYTHIA
calculations \cite{PYTHIA} above $\sqrt{s}$ = 10 GeV and extrapolated
to the individual thresholds, while the absolute strength of the cross
sections is fixed by the experimental data as described in Ref.
\cite{Cass01}. We here report on an extension of previous works
\cite{Cass01a,Cass97,Spieles} and include
explicitly the backward channels 'charm + anticharm meson $\rightarrow$
charmonia + meson' employing detailed balance in a more schematic
interaction model with a single parameter or matrix element $|M|^2$,
that is fixed by the $J/\Psi$ suppression data from the NA50
collaboration at SPS energies (cf. Ref. \cite{Brat03}).

Since the meson-meson dissociation
and backward reactions typically occur with low relative momenta
('comovers') it is legitimate to write the cross section for the
process $m_1 + m_2 \rightarrow m_3 +m_4$ as
\begin{equation}
\label{model}
 \sigma_{1+2 \rightarrow 3+4}(\sqrt{s}) = \ \frac{E_1 E_2 E_3
E_4}{s} \ |M_f|^2 \ \left( \frac{M_3+M_4}{\sqrt{s}} \right)^6 \
\frac{P_f}{P_i},
\end{equation}
where $E_i$ and $S_i$ denote the energy and spin of hadron $i$,
respectively, while $P_i$ and $P_f$ are the initial and final momenta for fixed invariant
energy  $\sqrt{s}$.  In
(\ref{model}) $|M_f|^2$ stands for the effective matrix element
squared,  which for the different 2-body channels is taken of the
form
\begin{eqnarray}
& |M_f|^2 = M_0^2 \hspace{1cm} & {\rm for} \ (\pi,\rho)+J/\Psi
   \rightarrow D+\bar{D} \label{mod}\\
& |M_f|^2 = 3 M_0^2 \hspace{1cm} & {\rm for} \ (\pi, \rho)+J/\Psi
   \rightarrow D^*+\bar{D}, D+\bar{D}^*, D^* + \bar{D}^* \nonumber\\
& |M_f|^2 = \frac{1}{3} M_0^2 \hspace{1cm} & {\rm for} \ (K,K^*)+J/\Psi
   \rightarrow D_s + \bar{D}, \bar{D}_s + D \nonumber \\
& |M_f|^2 =  M_0^2 \hspace{1cm} & {\rm for} \ (K,K^*)+J/\Psi
   \rightarrow D_s + \bar{D}^*, \bar{D}_s + D^*, D^*_s + \bar{D},
    \bar{D}^*_s + D, \bar{D}^*_s +D^* \nonumber
\label{mf}\end{eqnarray}
involving a single parameter $M_0^2$ (cf.
 \cite{Brat03} and Refs. therein). The relative factors of 3 in (\ref{mod}) are
guided by the sum rule studies in \cite{korean} which suggest that
the cross section is increased whenever a vector meson $D^*$ or
$\bar{D}^*$ appears in the final channel while another factor of
1/3 is introduced for each $s$ or $\bar{s}$ quark involved. The
factor $\left( {(M_3+M_4)}/{\sqrt{s}} \right)^6 $ in (\ref{model})
accounts for the suppression of binary channels with increasing
$\sqrt{s}$ and has been fitted to the experimental data for the
reactions $\pi + N \rightarrow \rho+N, \omega+N, \Phi+N, K^+
+\Lambda$ in Ref. \cite{CaKo}.
The advantage of the model introduced in (\ref{model}) is that
detailed balance for the binary reactions can be employed
strictly for each individual channel  and the role of the backward reactions
($J/\Psi$+meson formation by $D+\bar{D}$ flavor exchange) can be
explored without introducing any additional parameter once $M_0^2$
is fixed.

\begin{figure}[t]
\begin{center}
\begin{minipage}[l]{12 cm}
\epsfig{file=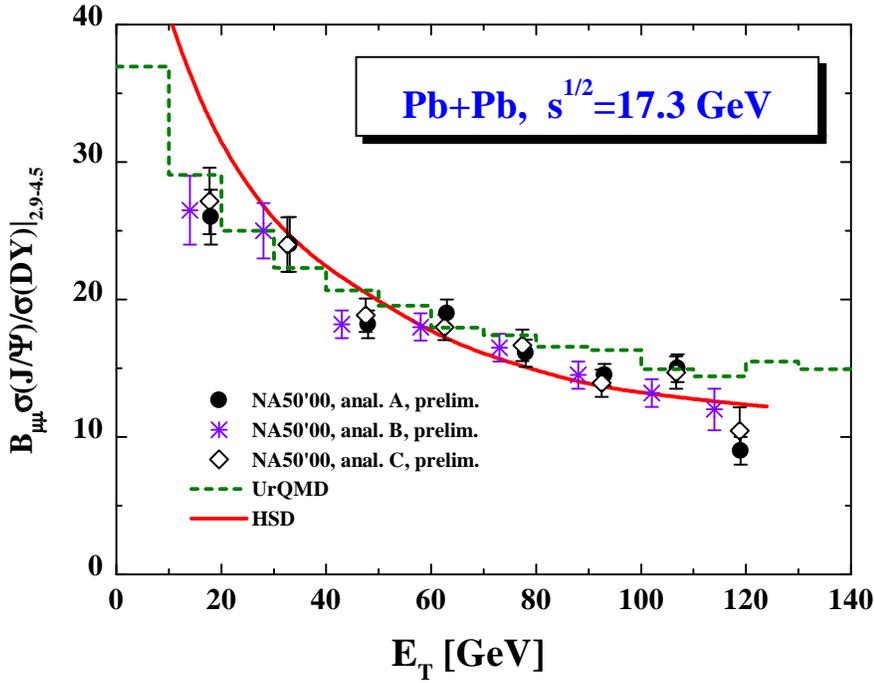,scale=0.75}
\end{minipage}
\phantom{a}\begin{minipage}[l]{5.8cm} \caption{The $J/\Psi$
suppression (in terms of the $\mu^+ \mu^-$ decay branch relative
to the Drell-Yan background from 2.9 -- 4.5 GeV invariant mass) as
a function of the transverse energy $E_T$ in $Pb~+~Pb$ collisions
at 160 A~GeV. The solid line shows the HSD result within the
comover absorption scenario presented in Section 2.3
\protect\cite{Brat03}. The different symbols stand for the NA50
data \protect\cite{NA50_QM02} from the year 2000 (ana\-lysis
A,B,C) while the dashed histogram is the UrQMD result from Ref.
\protect\cite{Spieles}.} \label{Fig2}
\end{minipage}
\end{center}
\end{figure}

We directly step on with the results for the charmonium
suppression and start with the system $Pb+Pb$ at 160 A~GeV to
demonstrate that the 'late' comover dissociation model
(\ref{model}) is approximately in line with the data of the NA50
Collaboration. The corresponding $J/\Psi$ suppression (in terms of
the $\mu^+ \mu^-$ decay branch relative to the Drell-Yan
background from 2.9 -- 4.5 GeV invariant mass) as a function of
the transverse energy $E_T$ in $Pb~+~Pb$ collisions at 160 A~GeV
is shown in Fig. 2. The solid line stands for the HSD result
within the  comover absorption scenario for the cross sections
defined by (\ref{model}) while the various data points reflect the
preliminary NA50 data from the year 2000 (analysis A,B,C).  A comparable
agreement with the data is also achieved within the UrQMD model
\cite{Spieles} (dashed histogram in Fig. \ref{Fig2}). We mention that
there might be alternative explanations for $J/\Psi$ suppression as
discussed in Refs. \cite{Satz99,Rappnew} and/or further dissociation
mechanism not considered here. However, for the purposes of the present
study it is sufficient to point out that the cross sections employed
here (cf. Figs. 6 and 7 in \cite{Brat03}) most likely are upper limits
and do not lead to a sizeable recreation of charmonia by $D+\bar{D}$
channels at SPS energies.

\begin{figure}[t]
\begin{center}
\begin{minipage}[l]{12 cm}
\epsfig{file=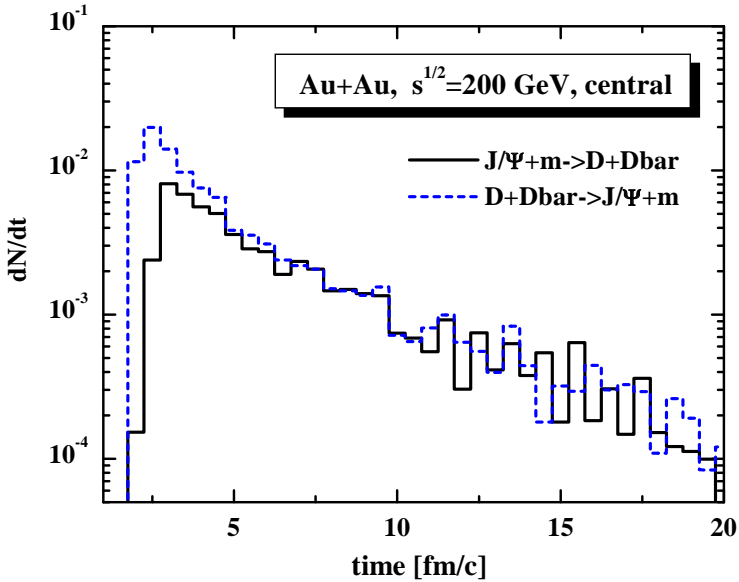,scale=0.55}
\end{minipage}
\phantom{a}\begin{minipage}[l]{5.8cm} \caption{The calculated rate
of $J/\Psi$ dissociation reactions with mesons (solid histogram)
for central $Au+Au$ collisions at $\sqrt{s}$ = 200 GeV in
comparison to the rate of backward reactions of open charm pairs to
$J/\Psi$ + meson (dashed histogram) according to the model
specified in Section 2.3 (taken from Ref. \protect\cite{Brat03}). } \label{bild11n}
\end{minipage}
\end{center}
\end{figure}

For central $Au+Au$ collisions at $\sqrt{s}$ = 200 GeV, however,
the multiplicity of open charm pairs should be  about 2 orders of
magnitude larger than at 160 A$\cdot$GeV (according to Fig. 1),
such that a much higher $J/\Psi$ reformation rate ($\sim
N_{c\bar{c}}^2$) is expected at RHIC energies (cf. Ref.
\cite{Rappnew}). In Fig. \ref{bild11n} we display the total
$J/\Psi$ comover absorption rate (solid histogram)  in comparison
to the $J/\Psi$ reformation rate (dashed histogram) as a function
of time in the center-of-mass frame for central Au+Au collisions
at $\sqrt{s}$ = 200 GeV.  The two rates become comparable for $t
\geq$ 4-5 fm/c and suggest that at the full RHIC energy of
$\sqrt{s}$ = 200 GeV the $J/\Psi$ comover dissociation is no
longer important since the charmonia dissociated in this channel
are approximately recreated in the backward channels. Accordingly,
the $J/\Psi$ dissociation at RHIC should be less pronounced  than
at SPS energies. Moreover, there is even a small excess of
$J/\Psi$ formation by $D+\bar{D}$ reactions in the first 2 fm/c.

\begin{figure}[t]
\begin{center}
\begin{minipage}[l]{12 cm}
\epsfig{file=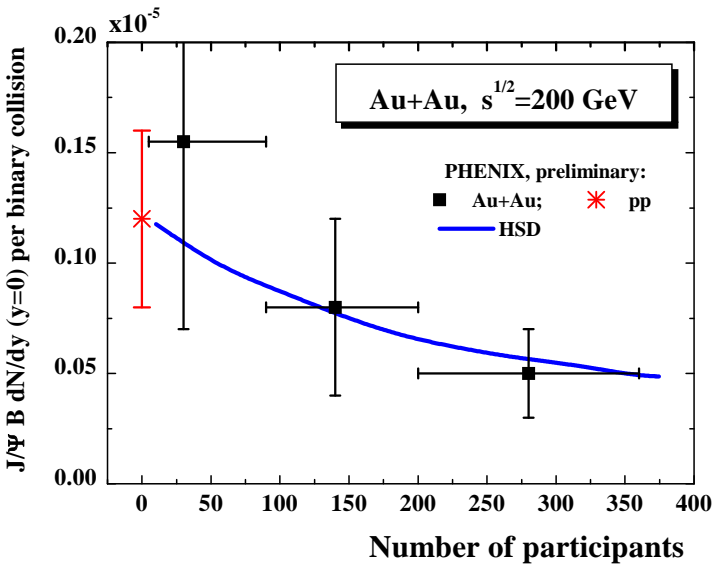,scale=0.60}
\end{minipage}
\phantom{a}\begin{minipage}[l]{5.8cm} \caption{The calculated
$J/\Psi$ multiplicity per binary collision -- multiplied by the
branching to dileptons --  as a function of the number of
participating nucleons $N_{part}$ in comparison to the preliminary
data from the PHENIX Collaboration \protect\cite{PHENIX2} for
$Au+Au$ and $pp$ reactions (taken from Ref. \protect\cite{Brat03}).
} \label{bild14n}
\end{minipage}
\end{center}
\end{figure}

The preliminary data of the PHENIX Collaboration \cite{PHENIX2}
allow for a first glance at the situation encountered in $Au+Au$
collisions at $\sqrt{s}$ = 200 GeV. In order to compare with the
preliminary data we have performed a rapidity cut $\Delta y \leq
2$ in the calculations. In Fig. \ref{bild14n} the $J/\Psi$
multiplicity per binary collision (times the branching ratio $B$)
is shown as a function of the number of participating nucleons in
comparison to the data at midrapidity. Since the statistics in
quite limited so far on the experimental side, no final conclusion
can presently be drawn, however, the data neither suggest a
dramatic enhancement of $J/\Psi$ production nor a complete
'melting' of the charmonia in the QGP phase.

\section{High $p_T$ hadron suppression}

Before coming to actual calculations for  deuteron-nucleus or
nucleus-nucleus collisions it is important to briefly explain the
concept of 'leading' and 'secondary' hadrons in the transport
approaches, since this separation is of central importance for the
results to be discussed below. To this aim we recall that in a
high energy nucleon-nucleon collision two (or more) color-neutral
strings are assumed to be formed, which phenomenologically
describe the low energy 'coherent' gluon dynamics by means of a
color electric field, which is stretched between the colored ends
of each single string. The latter string ends are defined by the
space-time coordinates of the constituents, i.e. a diquark and
quark for a 'baryonic' string or quark and antiquark for a
'mesonic' string. These constituents quarks (diquarks or
antiquarks) are denoted as 'leading' quarks that constitute the
'leading' pre-hadrons.

The time that is needed for the fragmentation of the strings and
for the hadronization of the fragments we denote as {\it formation
time} $\tau_f$.  For simplicity we assume (in HSD), that the
formation time is a constant $\tau_f$ in the rest frame of each
hadron and that it does not depend on the particle species.
 We recall, that due to time dilatation the formation
time $t_f$ in any reference frame is then proportional to the
Lorentz $\gamma$-factor, i.e.
       $ t_f=\gamma\cdot\tau_f$.
The size of $\tau_f$ can be estimated by the time that the
constituents of the hadrons (with velocity $c$) need to travel a
distance of a typical hadronic radius (0.5--0.8~fm).

We assume  that hadrons, whose constituent quarks and antiquarks
are created from the vacuum in the string fragmentation, do {\it
not interact} with the surrounding nuclear medium within their
formation time $t_f$. For the leading pre-hadrons, i.e. those involving
quarks (antiquarks) from the struck nucleons, we adopt a reduced
effective cross section $\sigma_{lead}$ during the formation time
$t_f$ and the full hadronic cross section later on.  Due to
time dilatation light particles emerging from the middle of the
string might escape the hadronic fireball without further
interaction if they carry a high momentum relative to the rest
frame of the fireball. However, the hadrons with transverse
momenta larger than $\sim$6 GeV/c predominantly stem from the
string ends \cite{CGC} and therefore can interact directly with a reduced
cross section (see below).

The relative quark counting factors of 2/3 or 1/3 for the
interactions of a 'leading' pre-hadron with a baryon or meson
mentioned above might appear arbitrary and simplistic. However,
this concept has been proven to work rather well for
nucleus-nucleus collisions from SPS to RHIC energies
\cite{Weber02,Brat03} as well as in hadron formation and
attenuation in deep inelastic lepton scattering off nuclei
\cite{Falter}. Especially the latter reactions are important to
understand the attenuation of hadrons with high (longitudinal)
momentum in ordinary cold nuclear matter. The studies in Ref.
\cite{Falter} have demonstrated that the dominant final state
interactions (FSI) of the hadrons with maximum momentum - as
measured by the HERMES Collaboration \cite{HERMES} - are
compatible with the concepts described above. This also holds for
antiproton production and attenuation in proton-nucleus collisions
at AGS energies \cite{AGS02}. Both independent studies point
towards a hadron formation time $\tau_f$ in the order of 0.4 to 0.8 fm/c.

As mentioned above, the question of relevance is the fraction of
leading pre-hadrons as a function of transverse momentum $p_T$ for
the different hadron species. This information is extracted from
PYTHIA calculations and displayed in Fig. 5 for $pp$ collisions at
$\sqrt{s}$ = 200 GeV. Here one notices slight differences between
pions and especially antibaryons, however, the fraction of leading
pre-hadrons increases  with $p_T$ and saturates above $\sim$ 6
GeV/c. Thus at high momenta the major fraction of 'hadrons' is of
leading quark, antiquark or diquark origin and - according to the
assumptions stated above - may interact without delay with the
partonic/hadronic environment.
\begin{figure}[t]
\begin{center}
\begin{minipage}[l]{12 cm}
\epsfig{file=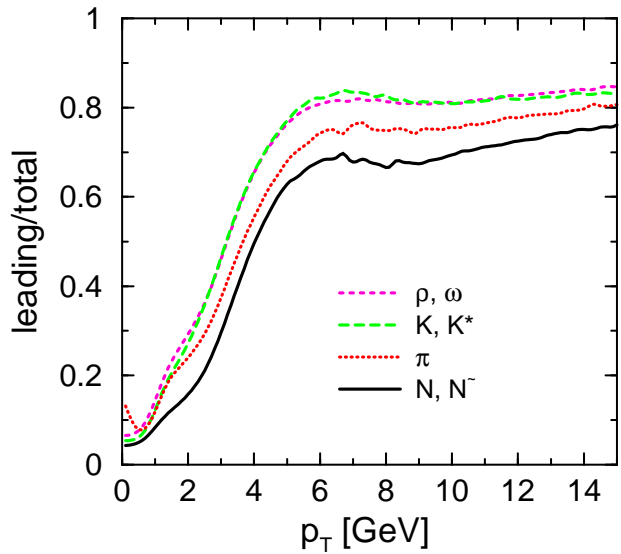,scale=0.45,angle=-90}
\end{minipage}
\phantom{a}\begin{minipage}[l]{6cm} \caption{The ratio of
'leading particles' to 'all produced particles' in
      $N+N$ collisions ($\sqrt{s}$=200 GeV) at midrapidity for
      different particle classes as a function of transverse
      momentum within the PYTHIA model (taken from Ref. \protect\cite{CGC}).}
\label{fig3}
\end{minipage}
\end{center}
\end{figure}

\subsection{{\it d+Au collisions at RHIC}}
As pointed out before (cf. Ref. \cite{Brat03PRL}) the slope of the
transverse mass spectra especially of kaons and antikaons at SPS
and RHIC energies is severely underestimated by the conventional
transport approaches HSD and UrQMD  when including only hadronic
reinteractions of secondary mesons. However, as known from the
experimental studies of Refs. \cite{Cronin1,Cronin2} an
enhancement of the transverse momentum cross section from $p+A$
collisions relative to scaled $pp$ collisions is already observed
at ISR energies. This 'Cronin effect' is presently not fully
understood in its details, but most likely related to an increase
of the average transverse momentum squared $<k_T^2>$ of the
partons in the nuclear medium prior to the 'hard' scattering
vertex. We speculate that such an enhancement of $<k_T^2>$ might
be related to initial semi-hard gluon radiation of the leading
parton in the medium, which is reduced in the vacuum due to color
neutrality in a small volume.  Since the microscopic mechanisms
are beyond the scope of our present analysis, we simulate this
effect in the transport approach by increasing the average
$<k_T^2>$ in the string fragmentation function with the number of
previous collisions $N_{prev}$ as
\begin{equation}
\label{kt}
<k_T^2> = <k_T^2>_{pp} (1+\alpha N_{prev}).
\end{equation}
The parameter $\alpha \approx 0.25-0.4$ is fixed in comparison to
the experimental data for d+Au collisions \cite{E1,E2} (cf. Fig.
6).

\begin{figure}[t]
\begin{center}
\begin{minipage}[l]{12 cm}
\epsfig{file=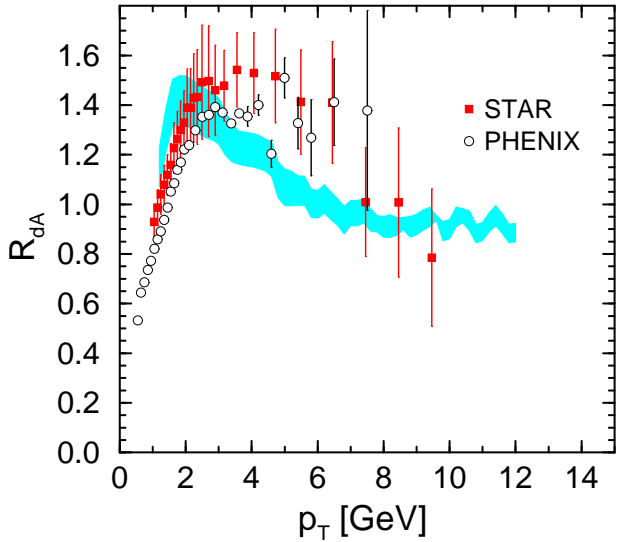,scale=0.45,angle=-90}
\end{minipage}
\phantom{a}\begin{minipage}[l]{5.8cm} \caption{The suppression
factor $R_{dA}$ (5) for minimum bias
      $d+Au$ collisions ($\sqrt{s}$=200 GeV) at midrapidity.
      Experimental data are from PHENIX \cite{E1} (open circles)
      and STAR \cite{E2} (filled squares),
      while the hatched band  shows the calculated
       charged hadron ratio (taken from Ref. \protect\cite{CGC}).} \label{fig4}
\end{minipage}
\end{center}
\end{figure}

A comparison of the calculated ratio
\begin{equation}
\label{ratio} R_{dA}(p_T) = \frac{1/N_{dA}^{event} d^2 N_{dA}/dy
dp_T}{<N_{coll}>/\sigma_{pp}^{inelas} d^2 \sigma_{pp}/dy dp_T}
\end{equation}
is shown in Fig. \ref{fig4}  with the respective data for charged
hadrons from Refs. \cite{E1,E2} for $\alpha = 0.25-0.4$ (hatched
band). In (\ref{ratio}) $<N_{coll}> \approx$ 8.5 denotes the
number of inelastic nucleon-nucleon collisions per event (for
minimum bias collisions), whereas $\sigma_{pp}^{inelas}$ denotes
the inelastic $pp$ cross section ($\approx$ 42 mb). We find that
the HSD calculations give a rise in the ratio (\ref{ratio}) for
small $p_T$ and a  decrease for $p_T \geq$ 2.5 GeV/c. The hatched
band gives an estimate for the systematic uncertainty in
(\ref{kt}) with respect to the strength of the Cronin effect that
carries over to studies for Au+Au reactions (see below). The
important issue in this context is that no dramatic absorption of
high $p_T$ hadrons is found in the calculations as well as in the
data.

\subsection{{\it Au+Au collisions at RHIC}}
\begin{figure}[t]
\begin{center}
\begin{minipage}[l]{12 cm}
\epsfig{file=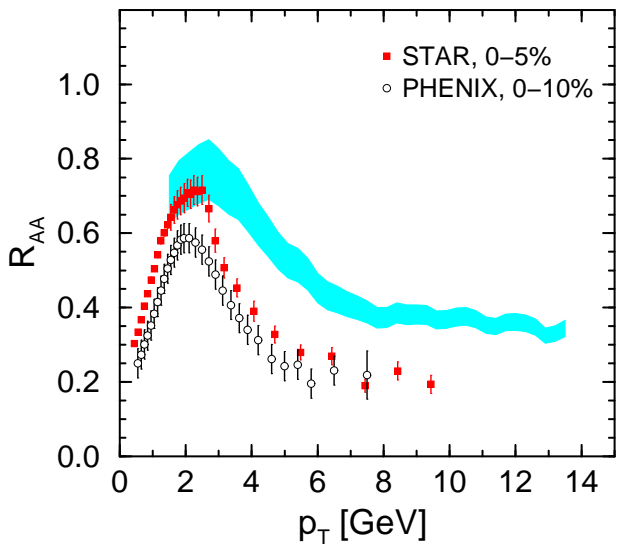,scale=0.45,angle=-90}
\end{minipage}
\phantom{a}\begin{minipage}[l]{5.8cm} \caption{The suppression
factor $R_{AA}$ (6) of charged hadrons
      at 5\% (or 10\%) central $Au+Au$ collisions ($\sqrt{s}$=200 GeV)
      at midrapidity for $\alpha$ = 0.25 to 0.4 (hatched band).
      The experimental data are from Refs. \cite{PHENIX,STAR} and show
      some additional suppression mechanism, which might be attributed
      to partonic interactions in a colored medium
      (taken from Ref. \protect\cite{CGC}).
} \label{fig5}
\end{minipage}
\end{center}
\end{figure}

We step on with 5\% central Au+Au collisions at $\sqrt{s}$ =
200 GeV. For Au+Au collisions we define the nuclear modification
factor in accordance with eq.~(\ref{ratio}) as
\begin{equation}
  \label{ratioAA}
  R_{AA}(p_T) = \frac{1/N_{AA}^{\rm event}\ d^2N_{AA}/dy dp_T}
  {\left<N_{\rm coll}\right>/\sigma_{pp}^{\rm inelas}\ d^2
    \sigma_{pp}/dy dp_T}\ .
\end{equation}
Fig.~\ref{fig5} shows a comparison of the calculations for
eq.~(\ref{ratioAA}) with the data for charged hadrons from Refs.
\cite{PHENIX,STAR}.
The surprising result in Fig. \ref{fig5} is that the ratio
$R_{AA}$ is roughly described in shape for the default HSD
parameters (formation time $\tau_F$ = 0.8 fm/c) within the
uncertainties of the 'Cronin' simulations. Only the absolute
magnitude of the high $p_T$ suppression seen experimentally is
slightly underestimated. In order to understand this result we
have decomposed the ratio $R_{AA}$ into contributions from
secondary pions, kaons and $\rho, \omega, K^*$ vector mesons
(before decays) as well as from leading pre-hadrons. We find that
formation time effects
 play a substantial role since heavier hadrons are
formed earlier than light pions in the cms frame at fixed
transverse momentum due to the lower Lorentz boost. However, the
total suppression by scatterings of formed hadrons is negligible
for $p_T \geq$ 6 GeV/c and only of relevance at lower $p_T$ for
the shape of $R_{AA}(p_T)$. Furthermore, the attenuation of the
leading pre-hadrons is roughly independent on $p_T$ and hadron
type and gives $R_{AA} \approx$ 0.1. Thus a large part of the
attenuation seen experimentally should be addressed to the
interactions of leading pre-hadrons with already formed hadrons.
Nevertheless, the experimental data indicate some additional
suppression mechanism.

\section{Conclusions}

Summarizing this contribution, we point out that strange hadron
production in central Au+Au (or Pb+Pb) collisions is quite well
described in the independent transport approaches HSD and UrQMD
\cite{Weber02}. The exception are the pion rapidity spectra at the
highest AGS energy and lower SPS energies, which are overestimated
by both models.  As a consequence the HSD and UrQMD transport
approaches underestimate the experimental maximum of the
$K^+/\pi^+$ ratio at $\sim$ 20-30 A$\cdot$GeV
\cite{Weber02,Horst}.

Furthermore, the transport calculations show that the charmonium
recreation by $D+\bar{D} \rightarrow J/\Psi + \ meson$ reactions
is comparable to the dissociation by 'comoving' mesons at RHIC
energies contrary to SPS energies. This leads to the final result
that the total $J/\Psi$ suppression as a function of centrality at
RHIC should be less than the suppression seen at SPS energies
where the 'comover' dissociation is substantial and the backward
channels play no role. Since the statistics is quite limited in
the present experiments for the $J/\Psi$ yield versus centrality,
no final conclusion can  be drawn so far; however, the available
data neither suggest a dramatic enhancement of $J/\Psi$ production
nor a complete 'melting' of the charmonia in a QGP phase
\cite{Brat03}.

Moreover, our transport calculations in comparison to experimental
data on high transverse momentum spectra from  d+Au and Au+Au
reactions show that pre-hadronic effects are responsible for both
the hardening of the spectra for low transverse momenta
\cite{Brat03c} (Cronin effect) as  well as the suppression of high
$p_T$ hadrons \cite{CGC}. The interactions of formed hadrons are
found to be negligible in central Au+Au collisions at $\sqrt{s}$ =
200 GeV for $p_T \geq$ 6 GeV/c and the large suppression seen
experimentally is attributed to a large extent to the interactions
of 'leading' pre-hadrons with the dense environment, which should
be partly of partonic nature in order to explain the large
attenuation seen in central Au+Au collisions.


\end{document}